\documentclass[3p,times]{elsarticle}

\usepackage{ecrc}

\volume{00}

\firstpage{1}

\journalname{Nuclear Physics A}

\usepackage{amssymb}

\biboptions{square,comma,numbers,sort&compress}

\usepackage[figuresright]{rotating}

\usepackage{graphicx}
\usepackage{dcolumn}
\usepackage{bm}
\usepackage{amssymb}
\usepackage{xspace}

\usepackage{hyperref}
\usepackage{amsmath}

\usepackage{color}

\newcommand{\pt}{\ensuremath{p_{\rm{T}}}\xspace}
\newcommand{\pp}{\ensuremath{\rm pp}\xspace}
\newcommand{\ppb}{p-Pb\xspace}
\newcommand{\pbpb}{Pb-Pb\xspace}
\newcommand{\nch}{\ensuremath{N_{\rm ch}}\xspace}
\newcommand{\nmpi}{\ensuremath{N_{\rm mpi}}\xspace}

\newcommand{\so}{\ensuremath{S_{\rm 0}}\xspace}
\newcommand{\meanpt}{\ensuremath{\langle p_{\rm T} \rangle}\xspace}

\begin{document}
\begin{frontmatter}


\dochead{}

\title{Mid-rapidity charged hadron transverse spherocity \\ in pp collisions simulated with Pythia}
\author{Eleazar Cuautle}
\author{Antonio Ortiz}
\author{Guy Pai\'c}

\address{Instituto de Ciencias Nucleares, Universidad Nacional Aut\'onoma de M\'exico. \\ Circuito exterior s/n, Ciudad Universitaria, Del. Coyoac\'an, C.P. 04510, M\'exico DF.}




\begin{abstract}

The \pp collisions have been studied for a long time, however, there are still some effects which are not completely understood, such as the long range angular correlations and the flow patterns in high multiplicity events, which were recently discovered at the LHC. In a recent work it was demonstrated that in Pythia 8, multi-parton interactions and color reconnection can give some of the observed effects similar to the collective flow well known from heavy-ion collisions. Now using the same model, a study based on mid-rapidity charged hadron transverse spherocity is presented. The main purpose of this work is to show that a differential study combining multiplicity and event shapes opens the possibility to understand better the features of data, specially at high multiplicity.

\end{abstract}

\begin{keyword}
Multi-parton interactions, flow-like behavior, event shapes, \pp collisions.



\end{keyword}

\end{frontmatter}




\section{Introduction}
\label{intro}

Recent results at the LHC have uncovered hitherto unknown features in high multiplicity events.  Perhaps the most
interesting result is the discovery of sQGP-like (ridge-like and flow-like behavior) features in small systems like those created in pp and p-Pb collisions~\cite{Khachatryan:2010gv,CMS:2012qk,Abelev:2012ola,ABELEV:2013wsa,Abelev:2013haa,OrtizVelasquez2014146}. The origin of such effects is still far from being understood, for
instance, it has been shown that hydro calculations, where the formation of a hot and dense QCD medium is implicitly assumed, can describe qualitatively many features of data~\cite{PhysRevC.88.014903,PhysRevLett.111.172303,PhysRevLett.113.252301}. But, other approaches suggest that the phenomenon can be produced by initial state effects~\cite{Ma:2014pva,Schenke:2015aqa,Dumitru:2010iy}. For instance, Pythia 8.180~\cite{Sjostrand:2007gs} gives an effect reminiscent of the collective flow well known from heavy-ion collisions~\cite{Ortiz:2013yxa}. This flow-like behavior is attributed to multi-parton interactions (MPI) and color reconnection (CR)~\cite{Sjostrand:2013cya}.

The impression that the understanding of the initial stages of the heavy ion collisions is
reasonably known contrasts with the opinion of a representative
number of physicists who emphasize the fact that the understanding of 
processes transforming the initial stage into a hydrodynamical final
state is still incomplete and requires further theoretical and
experimental  knowledge~\cite{Antinori:2014xma}.

\begin{figure}
\begin{center}
\resizebox{0.5\textwidth}{!}{%
  \includegraphics{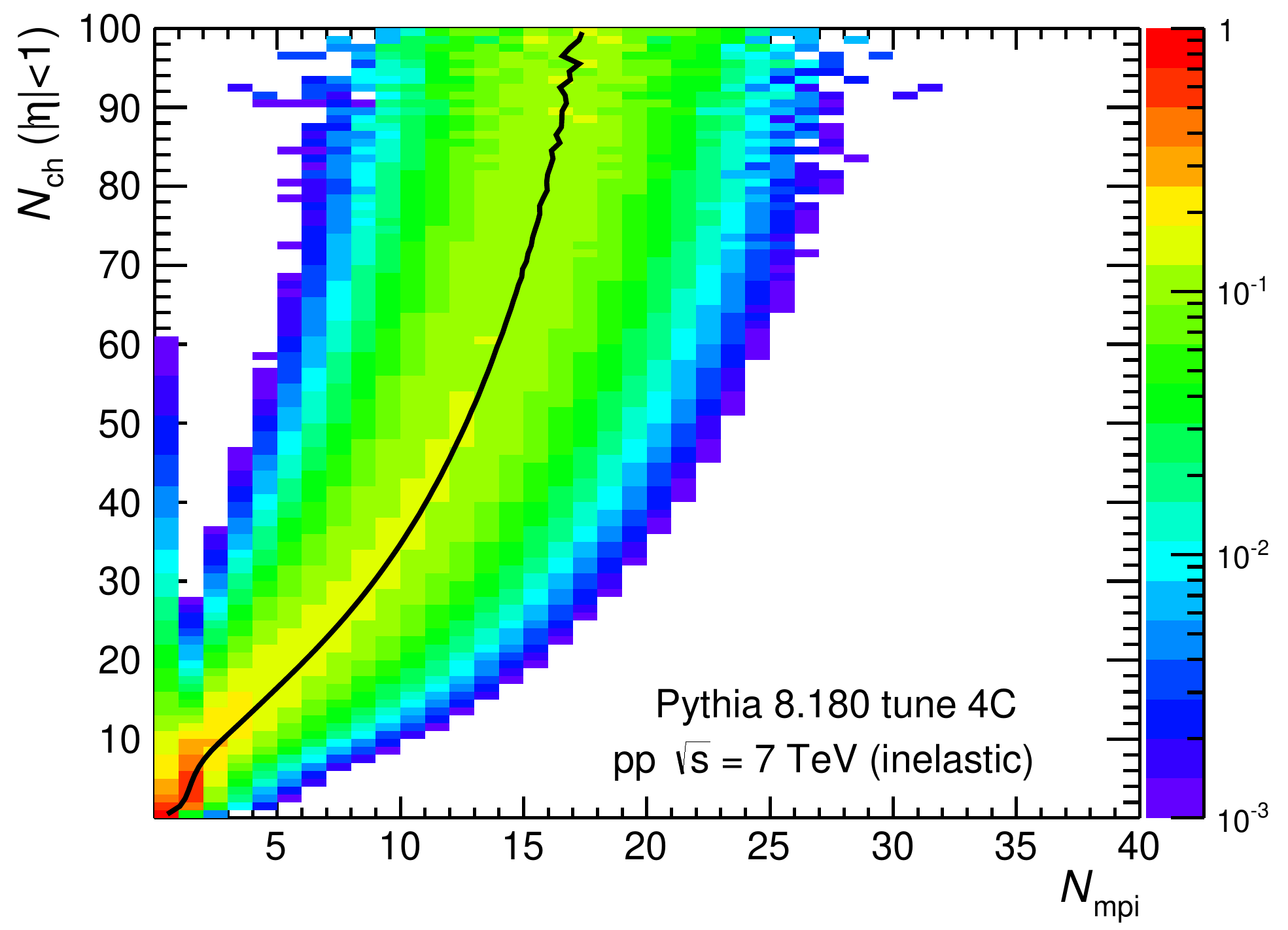}
}
\caption{(Color online). Mid-rapidity ($|\eta|<1$) charged hadron multiplicity as a function of the number of multi-partron interactions for \pp collisions at $\sqrt{s}=$ 7 TeV simulated with Pythia 8.180 tune 4C. Each multiplicity interval is normalized to one. The solid line illustrates the average number of MPI in a given multiplicity interval.}
\label{fig:1}       
\end{center}
\end{figure}

In this work the results of an event shape analysis, specifically, a selection on transverse spherocity of mid-rapidity charged hadrons, applied to events generated with Pythia 8.180 are presented. The goal of the study is to show that the use of event shapes allows to extract more information from data. Namely, this technique opens the possibility to isolate jetty-like (high \pt jets) and isotropic (large number of low $Q^{2}$ partonic scatterings) events~\cite{Cuautle:2014yda}. It is worth to notice that the procedure is inverse to the usual approach. Namely, in the majority of cases one tries to apply models, which successfully describe heavy ion collision data, to smaller system,
sometimes forgetting that the premises valid in large systems are not
at all satisfied in smaller ones. For example, the requirement that the
Knudsen number  $K= \lambda/R$, where $\lambda$ is the mean free path and $R$ is
the dimension of the system, must be small  to allow for a rapid
equilibration; is not really warranted in small systems or similarly, for
particles that exhibit small interaction cross section with the medium. 

With the exception of the anisotropic flow measurements~\cite{PhysRevC.90.054901}, where the so-called non-flow contributions are well under control, most of the observables used to study the properties of the hot and dense QCD matter contain a mixture of contributions from the different components present in the collisions i.e., soft and hard QCD processes. For example, as illustrated in this work~\cite{Veldhoen:2012ge}, the identified hadron production measured in central \pbpb collisions is completely different in the jet region and outside the jet peak (bulk)\footnote{A similar result has been obtained from the analysis of  \ppb data using jet reconstruction and strange hadrons~\cite{Zimmermann:2015npa}.}. However, the low \pt ($<3$ GeV/$c$) part of the inclusive spectra is used to extract expansion velocity and the temperature at the kinetic freeze-out of the system~\cite{PhysRevLett.109.252301}.  In this context, it is argued
that the implementation of an event shape analysis would allow for a better
understanding of non-radial flow effects.

\section{Transverse spherocity}

In the present work, the mid-rapidity charged hadron transverse spherocity, \so, is used  to characterize the events through the geometrical distribution of the  \pt's of the charged hadrons,  which is by definition collinear and infrared safe. The restriction to
the transverse plane avoids  the  bias   from  the  boost   along  the
beam axis~\cite{Banfi:2010xy}. It is defined for a unit transverse vector $\mathbf{\hat{n}}$ which minimizes the ratio below:

\begin{equation}
S_{\rm 0} = \frac{\pi^{2}}{4} \left(  \frac{\sum_{i} |{{\overrightarrow{\pt}}}_{i} \times \mathbf{\hat{n}}|}{\sum_{i} {\pt}_{i}}  \right)^{2}.
\end{equation}

By construction, the limits of the variable are related to specific configurations in the transverse plane
\begin{displaymath}
S_{\rm 0} = \left\lbrace    \begin{array}{ll}
0 & \textrm{``pencil-like'' limit (hard events)} \\
1 & \textrm{``isotropic'' limit (soft events)}
\end{array} \right. \,.
\end{displaymath}
\noindent
In this study, inelastic \pp collisions were generated with Pythia 8.180 tune 4C, this tune describes qualitatively many features of the LHC \pp data~\cite{Corke:2010yf}. The event shape was computed considering only primary charged particles at mid-rapidity ($|\eta|<$1) and in the transverse momentum interval $0.15< \pt < 10$ GeV/$c$. Transverse spherocity was only defined for events with more than two hadrons. Different observables like jet production and identified hadron production were studied at mid-rapidity for different \so and multiplicity (\nch) intervals.

Event shapes studied by experiments at LHC, e.g. ATLAS~\cite{Aad:2012fza} and ALICE~\cite{Abelev:2012sk}, have shown an interesting result: a good agreement between data and models is observed for the average event shape, while the event shape distributions exhibit large discrepancies. This is a very important message: the average measurements do not contain enough information, hence, care needs to be taken when extracting physics from models.  For example, in the concrete case of the event generators reported in these references~\cite{Aad:2012fza, Abelev:2012sk}, they overestimate significantly the contribution of back-to-back jets events and underestimate the contribution of isotropic events at high multiplicity.

\begin{figure*}
\begin{center}
\resizebox{0.85\textwidth}{!}{%
  \includegraphics{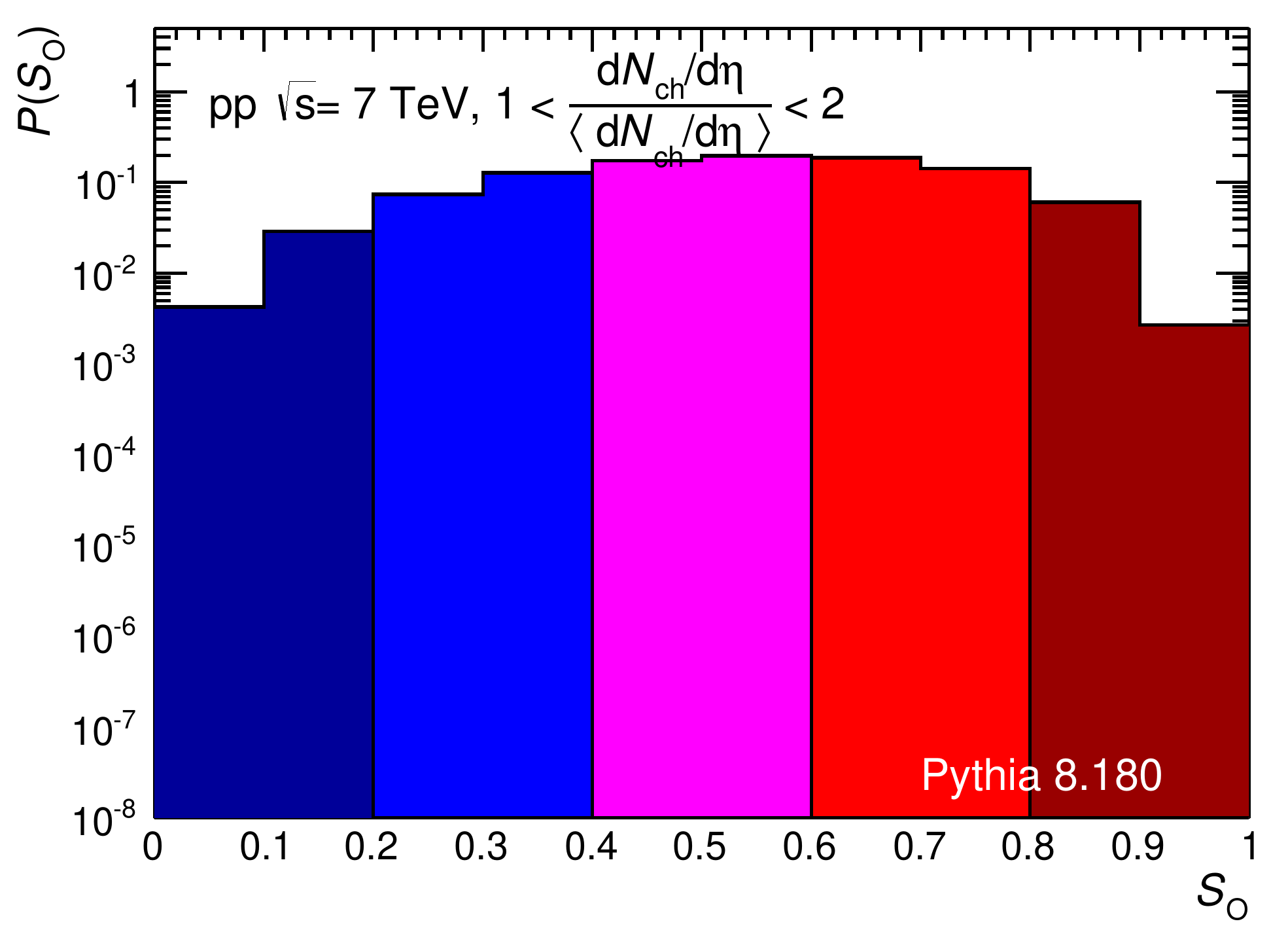}
  \includegraphics{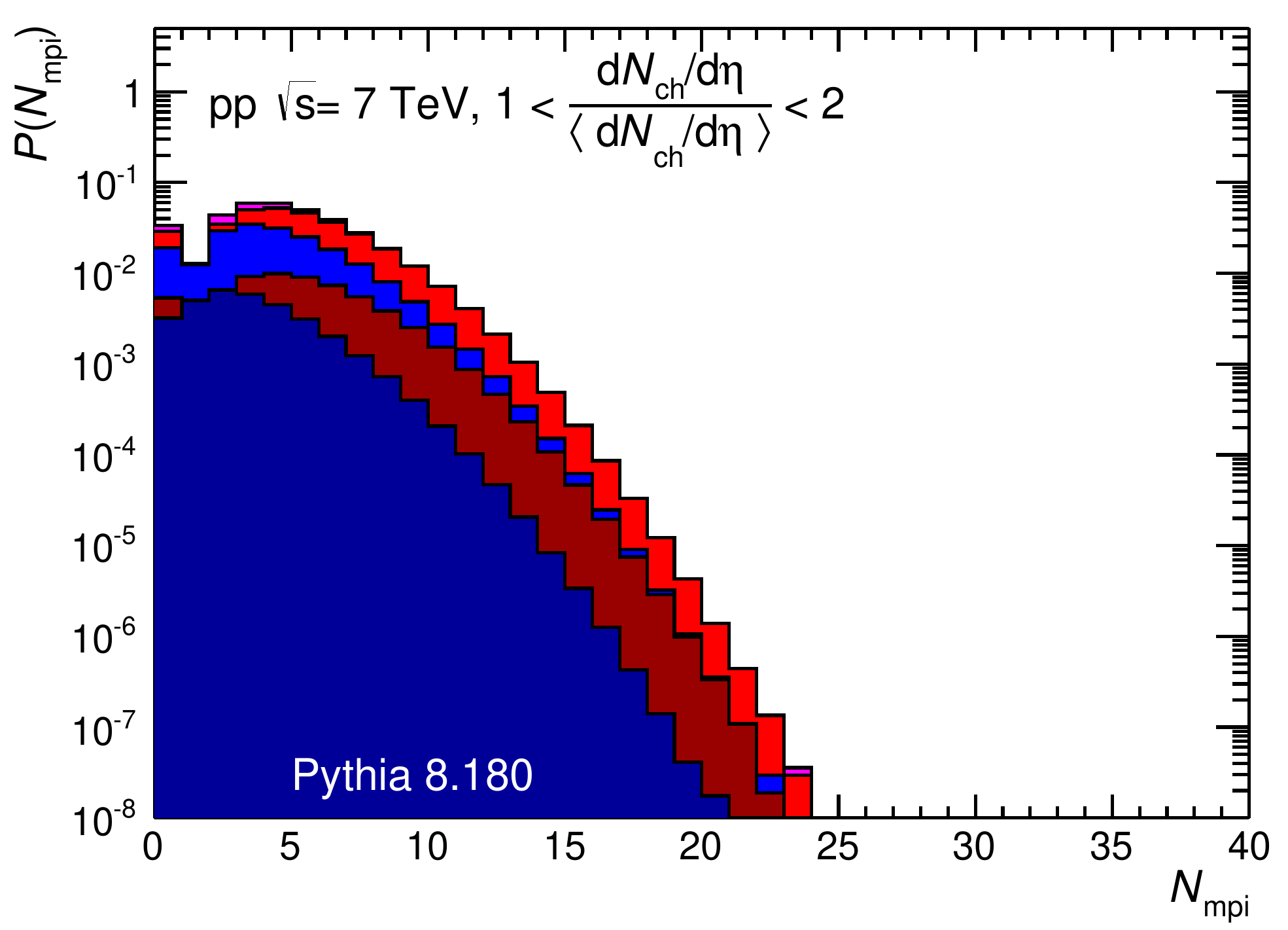}
}
\resizebox{0.85\textwidth}{!}{%
  \includegraphics{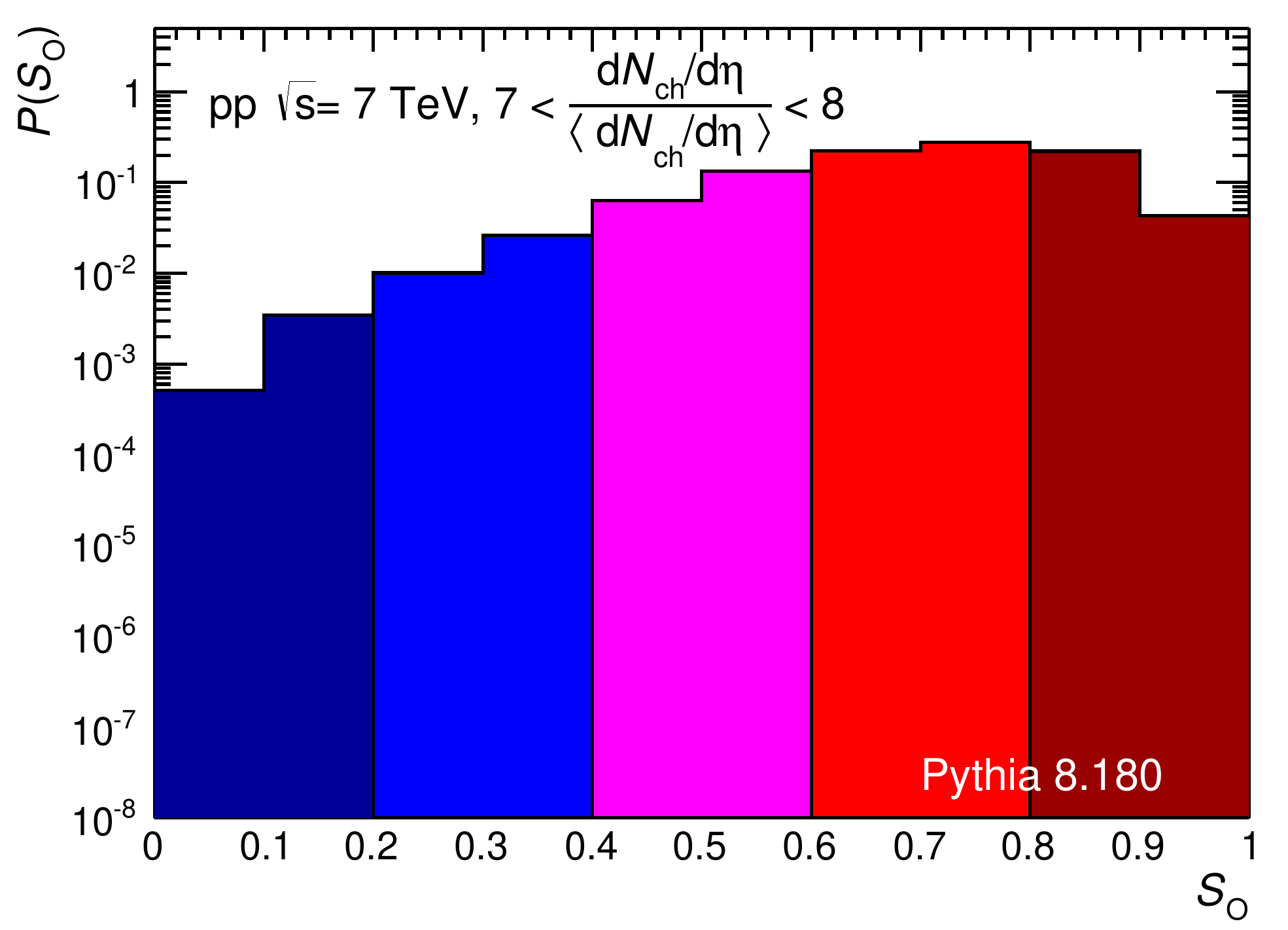}
  \includegraphics{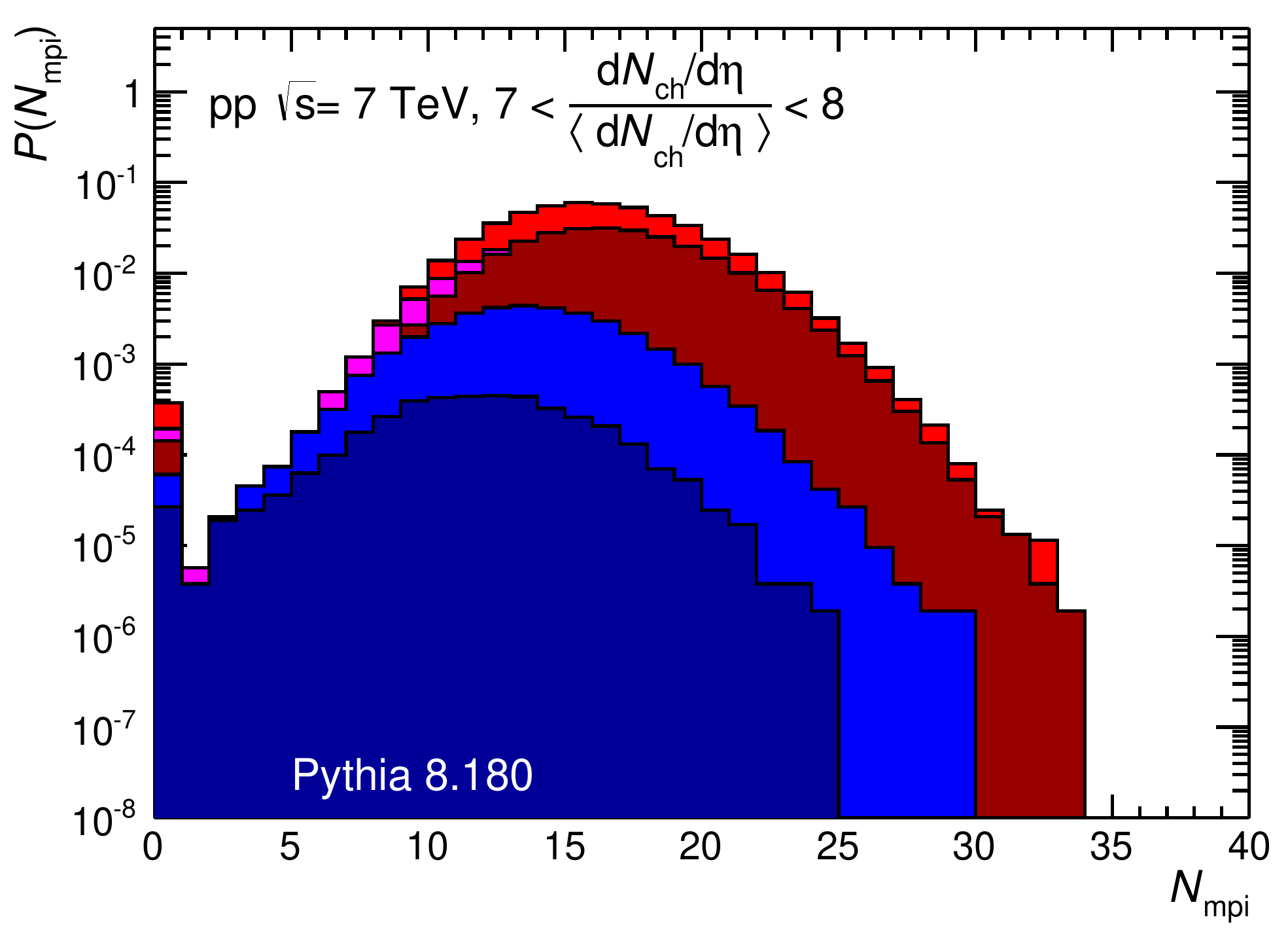}  
}
\vspace*{0.5cm}       
\caption{(Color online). Transverse spherocity distributions for $\sqrt{s}=$ 7 TeV \pp collisions producing low (upper left) and high (bottom left) mid-rapidity charged hadron multiplicity. The corresponding \nmpi distributions are shown in the right panels.}
\label{fig:2}       
\end{center}
\end{figure*}

\section{Multi-parton interactions, event multiplicity and transverse spherocity}
 
 MPI are a theoretically ~\cite{PhysRevD.36.2019} expected phenomenon, their effects
have been observed by several experiments~\cite{Zeus:2008, CDF:1997, D0:2010}. MPI in Pythia are crucial for
the description of observables like multiplicity distributions, underlying event,
correlations of the average transverse sphericity with multiplicity, and, together
with color reconnection, to produce flow-like patterns.

Figure~\ref{fig:1} shows the correlation between mid-rapidity charged hadron multiplicity and the number of multi-parton interactions (\nmpi) for \pp collisions at $\sqrt{s}=$ 7 TeV. Two features are observed. First, the width of the distribution for multi-parton interactions prevents using multiplicity alone as a selective parameter. Namely, in a given multiplicity interval there are events emanating from a very different number of MPI, hence of very different nature. The second feature is the tendency of saturation which is observed at high multiplicity. ALICE has measured the number of independent sources of particle production as a function of the event multiplicity using an approach based on two-particle azimuthal correlations, and it has reported a saturation effect at high multiplicity~\cite{ALICEwpazC}.

  The left panels of Fig.~\ref{fig:2} show the transverse spherocity spectra for \pp
collisions at 7 TeV for low  and high 
multiplicity events corresponding to values of ${\rm d}\nch/{\rm
d}\eta$ equal to one and seven times the average value obtained for minimum bias
(MB) \pp collisions, respectively. The
right panels show the multi-parton interaction distributions for both multiplicity
classes. A depletion of the low \so part in the high multiplicity events compared
with the low multiplicities is observed. The
phenomenon is accompanied by a much higher number of multi-parton interaction,
namely, for isotropic events a greater average \nmpi is obtained than for jetty-like
events. These results suggest the use of event shapes for a better
selectivity  of the events with a certain number of  MPI.

\section{Results}
\subsection{Jet production as a function of multiplicity and transverse spherocity}

As discussed earlier, event shapes are tools which allow the classification of the events according to their jet content. To probe the tool, this section shows the results of applying a jet finder to samples of events, where a pre-selection based on transverse spherocity was implemented. The jet finder Fast jet 3.0.6~\cite{Cacciari:2011ma} has been used, that implements the anti-$k_{\rm T}$ algorithm with a jet radius of 0.4. The minimum jet \pt was set to 10 GeV/$c$.

\begin{figure*}
\begin{center}
\resizebox{0.85\textwidth}{!}{%
  \includegraphics{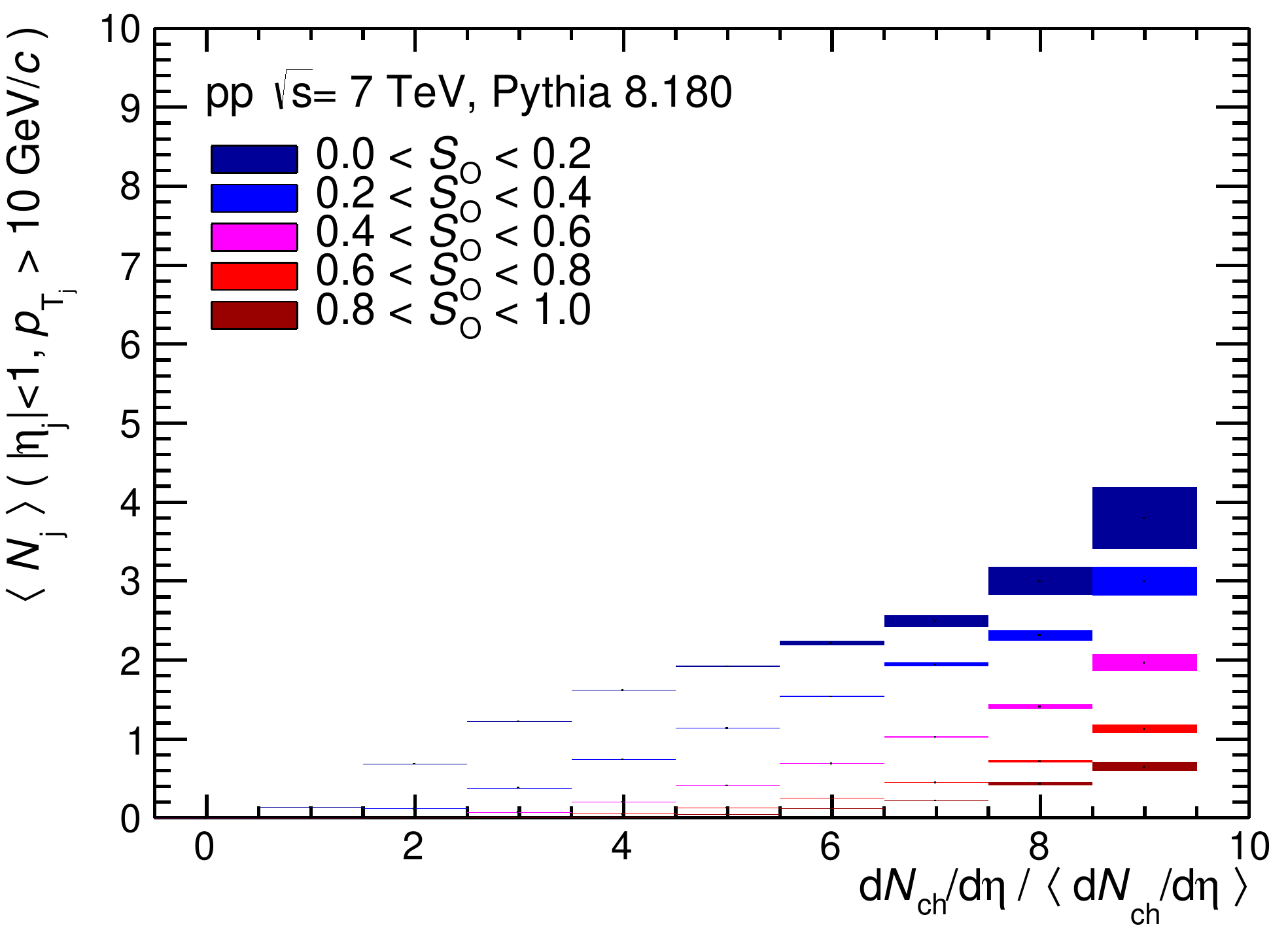}
  \includegraphics{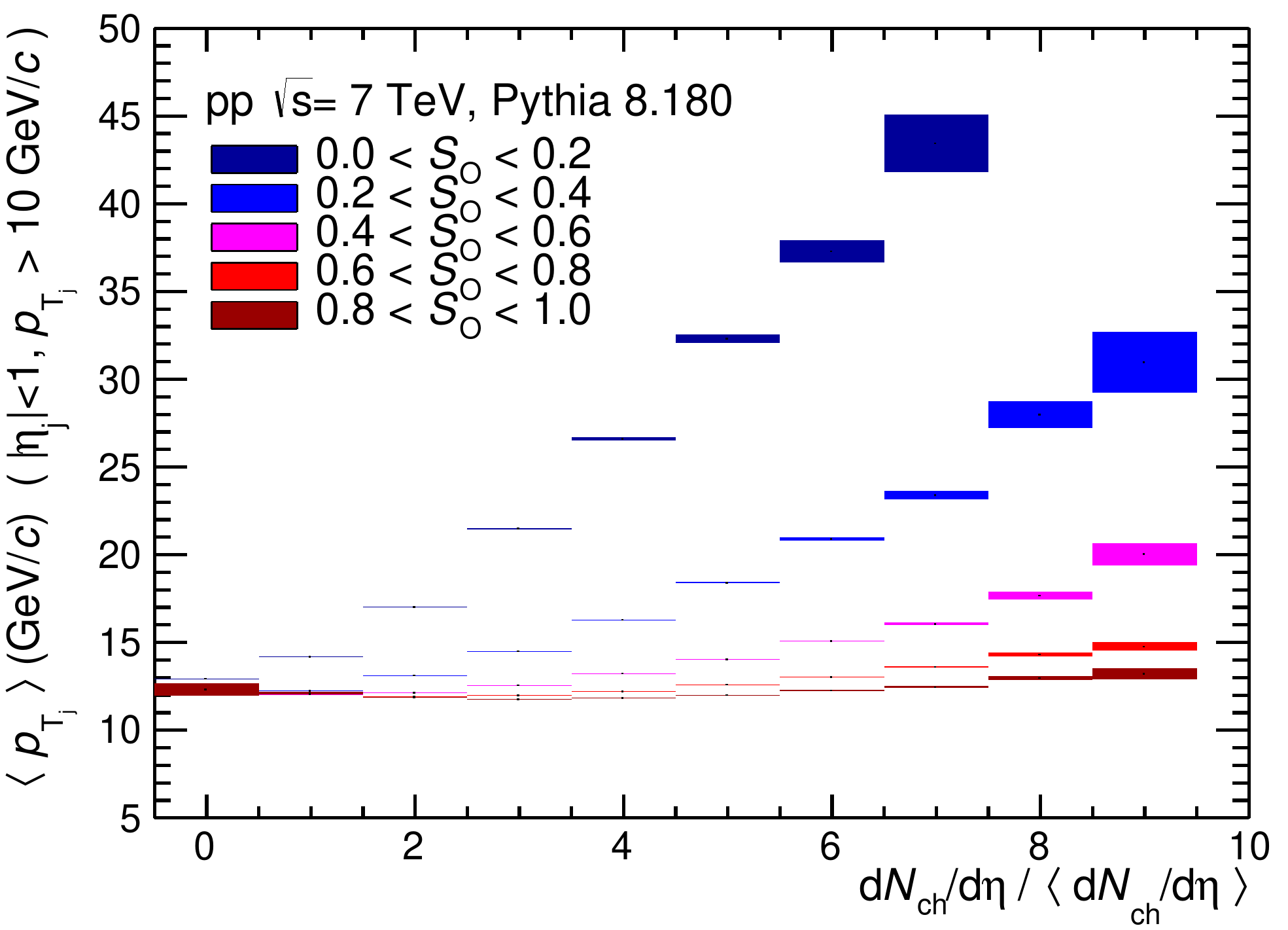}
}
\vspace*{0.5cm}       
\caption{(Color online). Average number of jets, reconstructed with Fast jet 3.0.6, as a function of the event multiplicity is shown in the left panel. The average jet \pt as a function of the event multiplicity is shown in the right panel. Multiplicity is normalized to the average \nch obtained for MB events.}
\label{fig:3}       
\end{center}
\end{figure*}

The left panel of Fig.~\ref{fig:3} shows the average number of jets as a function of the event multiplicity. Results are presented for different \so intervals. As expected from transverse spherocity definition, for isotropic events ($\so>0.8$) the average number of jets is below 1. This number increases when reducing transverse spherocity. Overall, the jet production rises with the event multiplicity, but the largest increase is seen in jetty-like events ($\so<0.2$). A complementary study is presented in the right panel, where the average jet \pt is plotted as a function of multiplicity. For isotropic events, the small fraction of jets which survives after the \so cut, exhibits an average \pt which is flat and close to 10 GeV/$c$. On the contrary, a strong dependence is found for jetty-like events where at high multiplicity the average \pt is above  30 GeV/$c$.  

\begin{figure*}
\begin{center}
\resizebox{0.85\textwidth}{!}{%
  \includegraphics{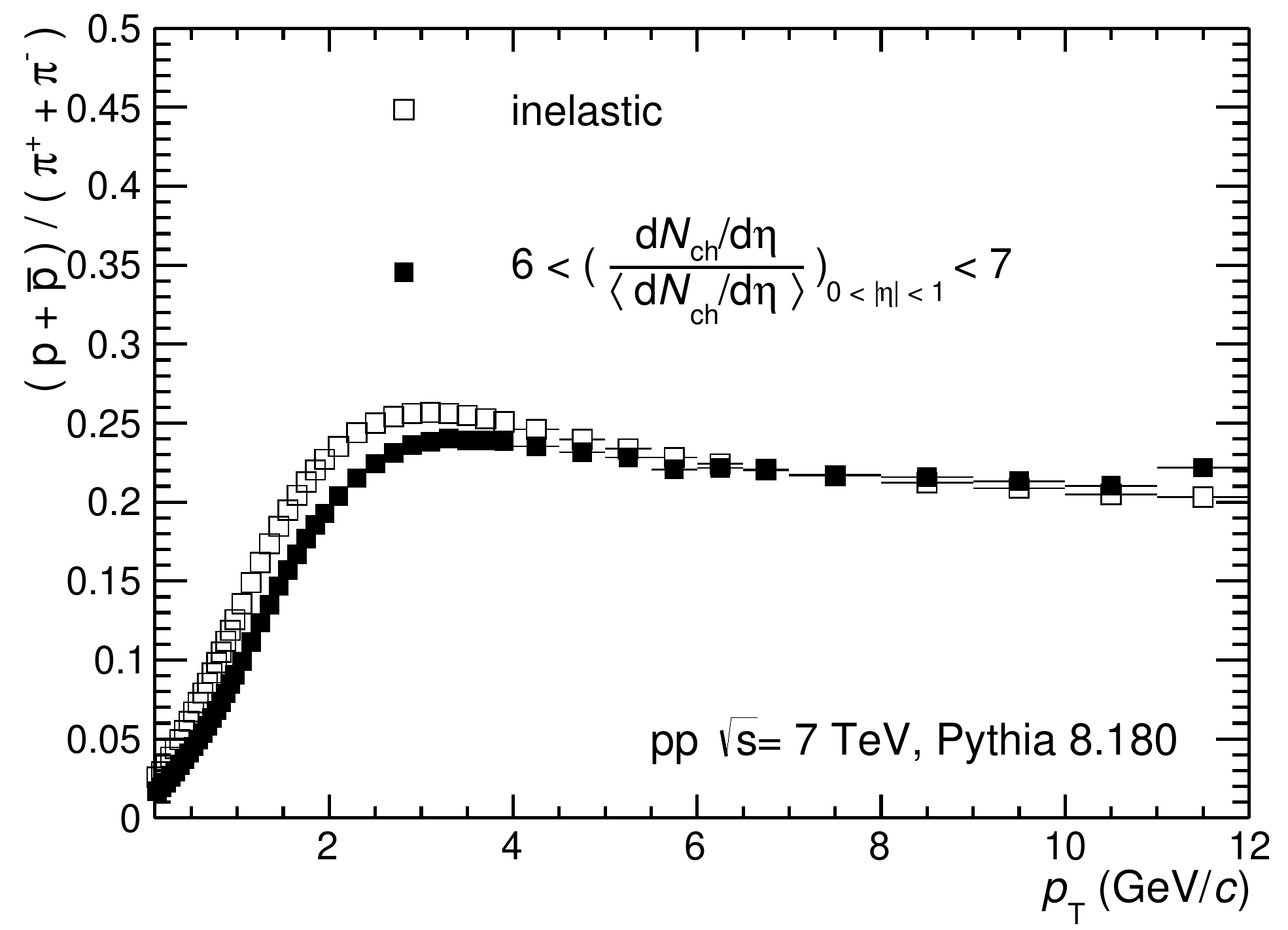}
  \includegraphics{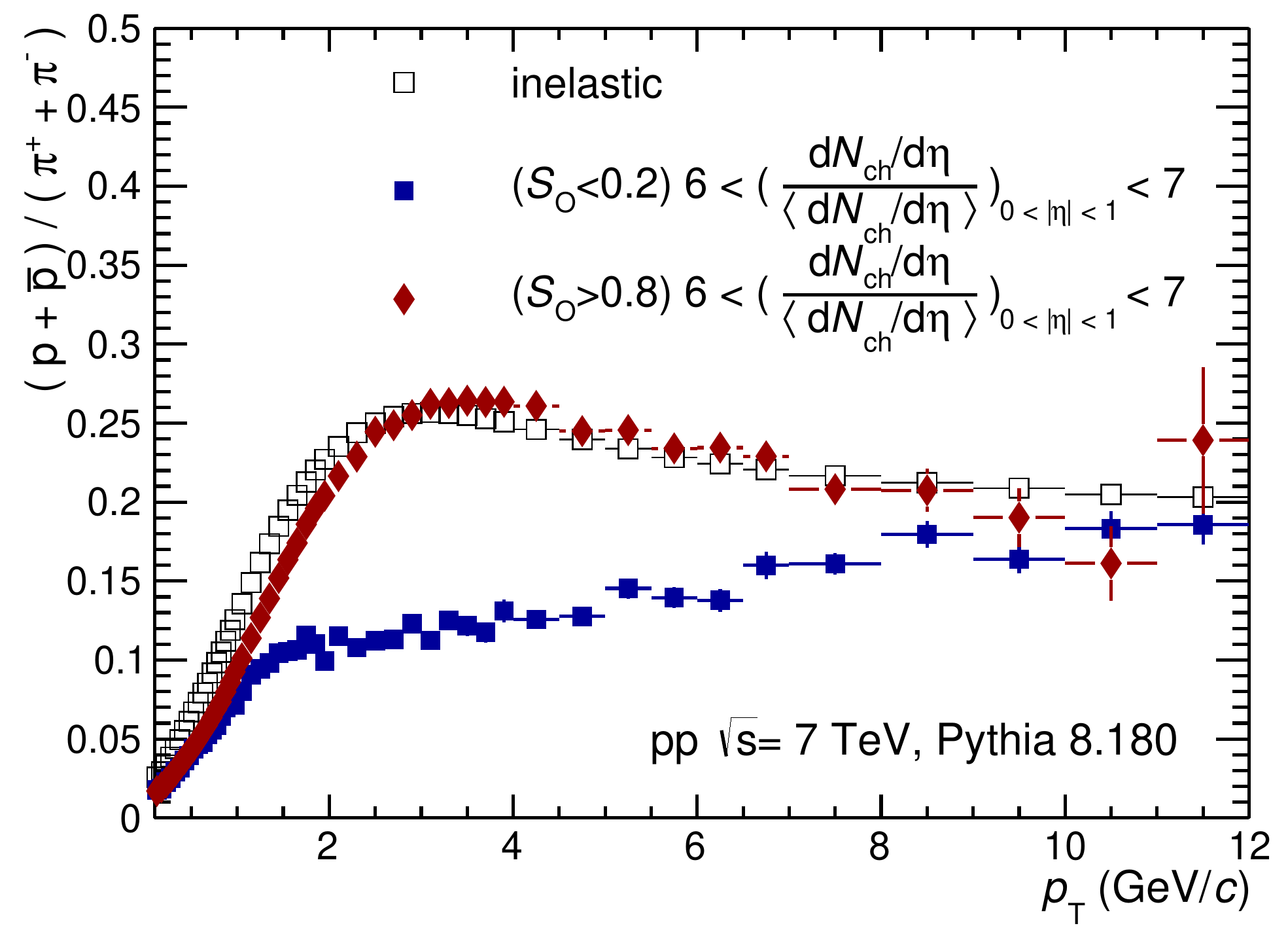}
}
\vspace*{0.5cm}       
\caption{(Color online). Proton yield normalized to pion yield as a function of transverse momentum. The result for MB is compared to that for high multiplicity events (left). Also a comparison with the cases where an additional selection on transverse spherocity is implemented (right).}
\label{fig:4}       
\end{center}
\end{figure*}

\begin{figure*}
\begin{center}
\resizebox{0.98\textwidth}{!}{%
  \includegraphics{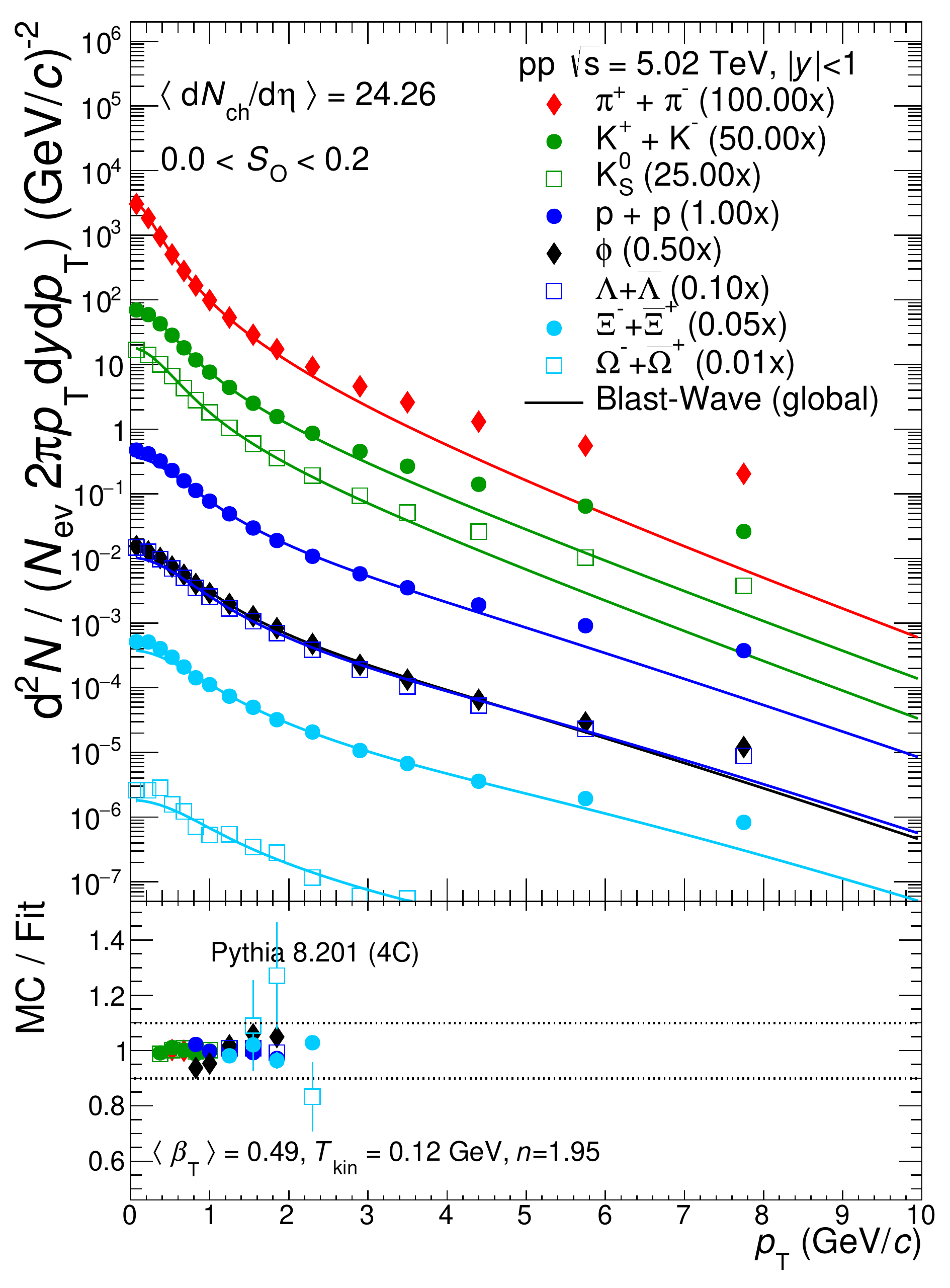}
  \includegraphics{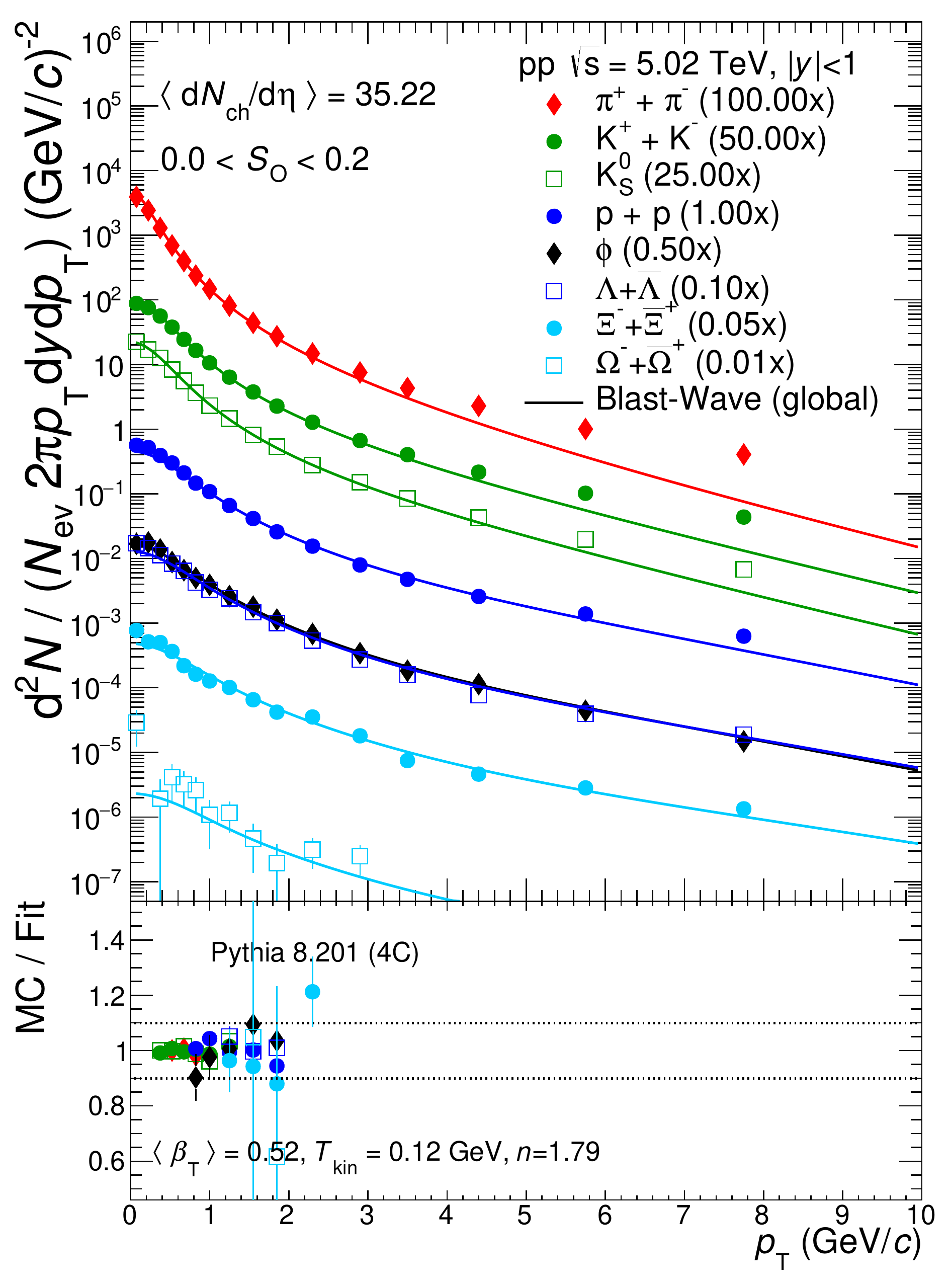}
  \includegraphics{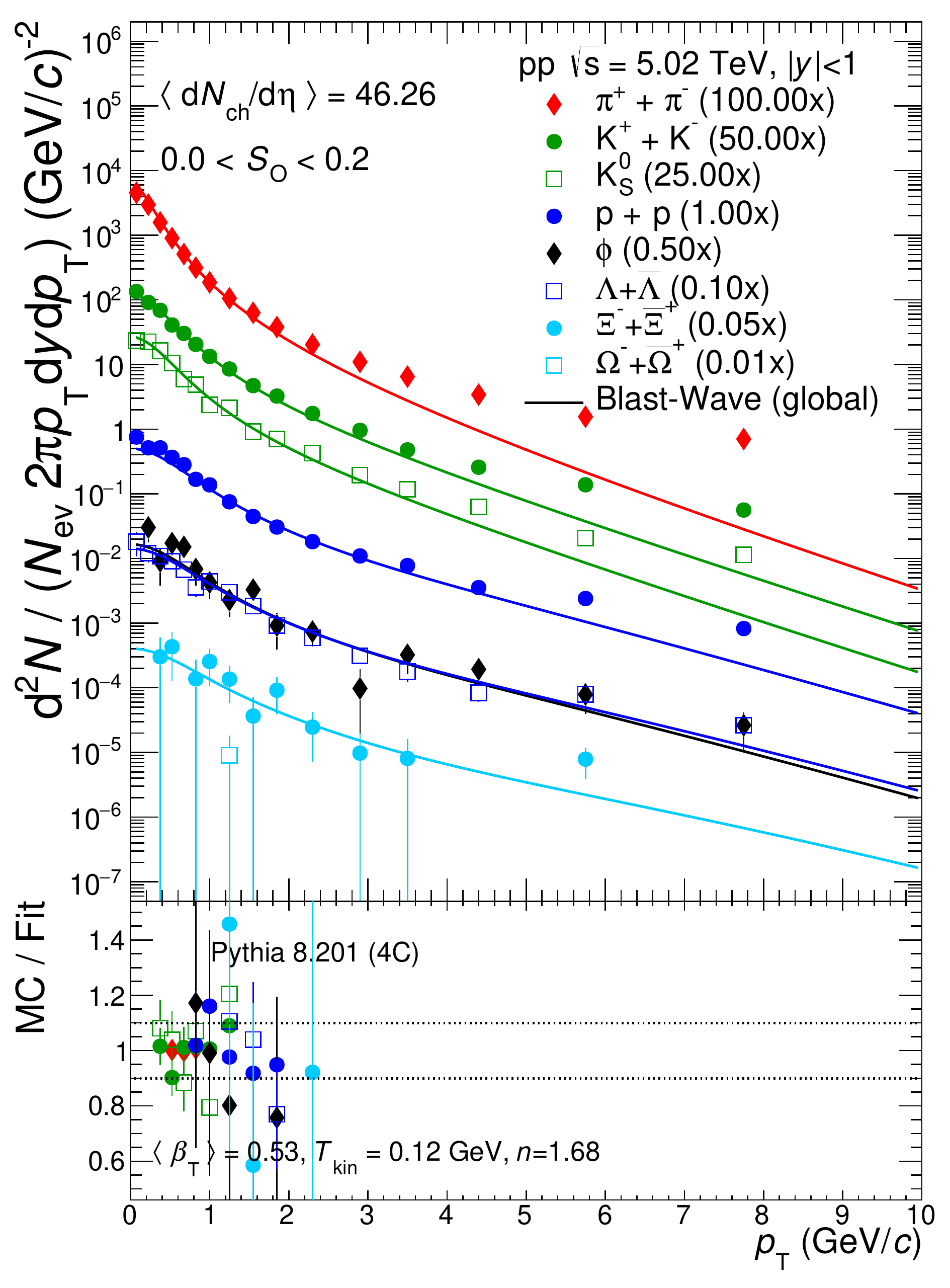}
}
\resizebox{0.98\textwidth}{!}{%
  \includegraphics{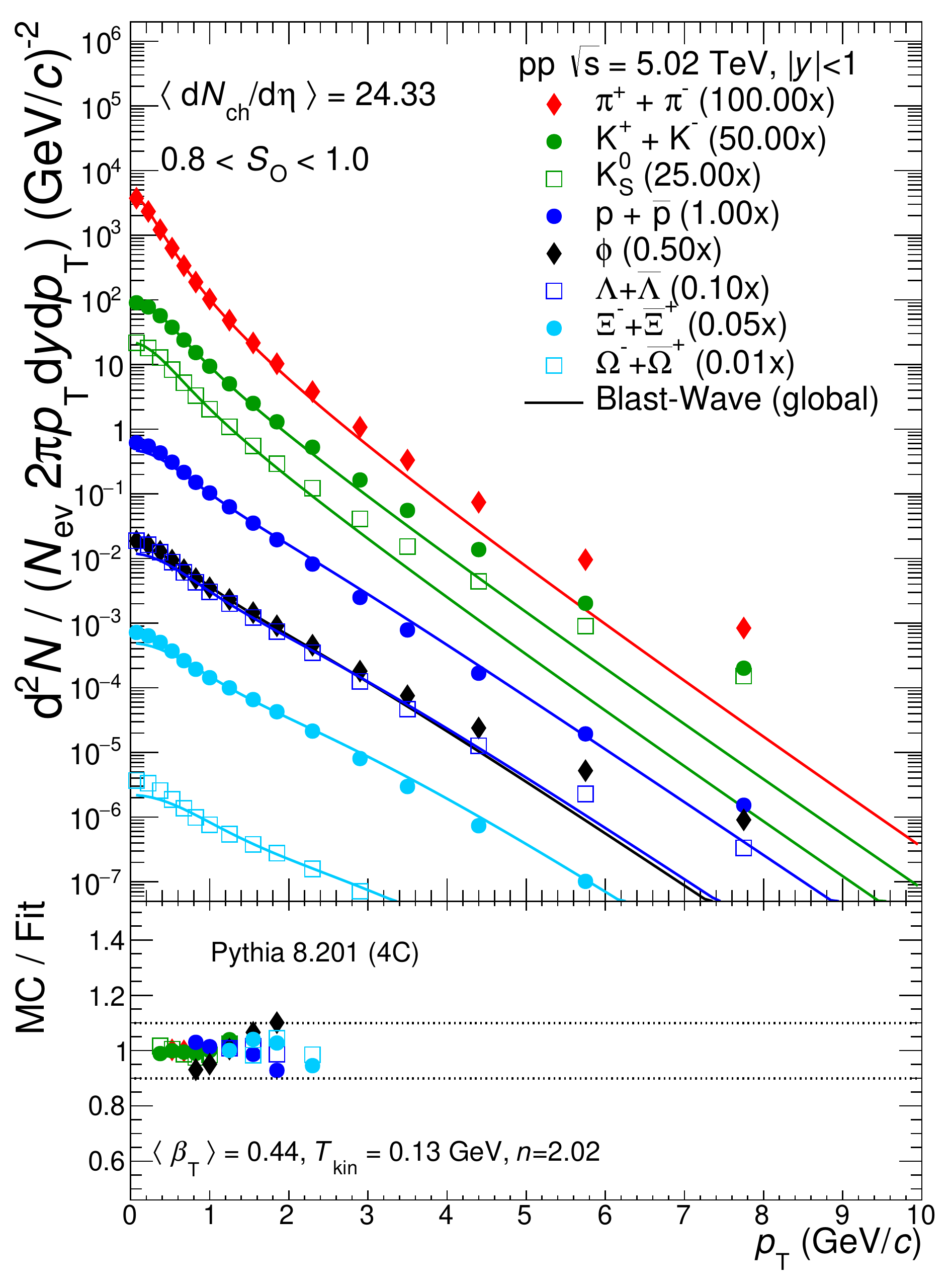}
  \includegraphics{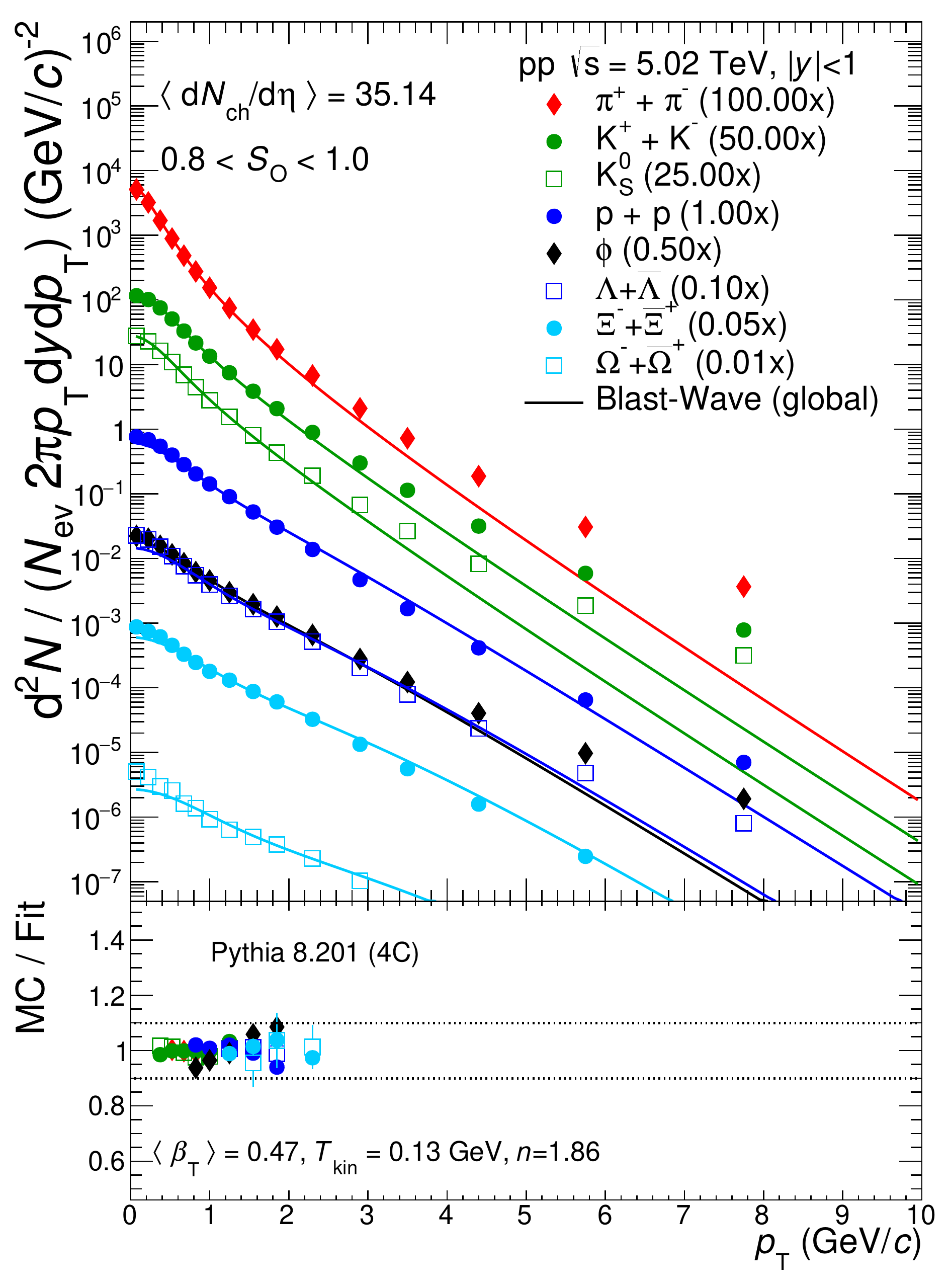}
  \includegraphics{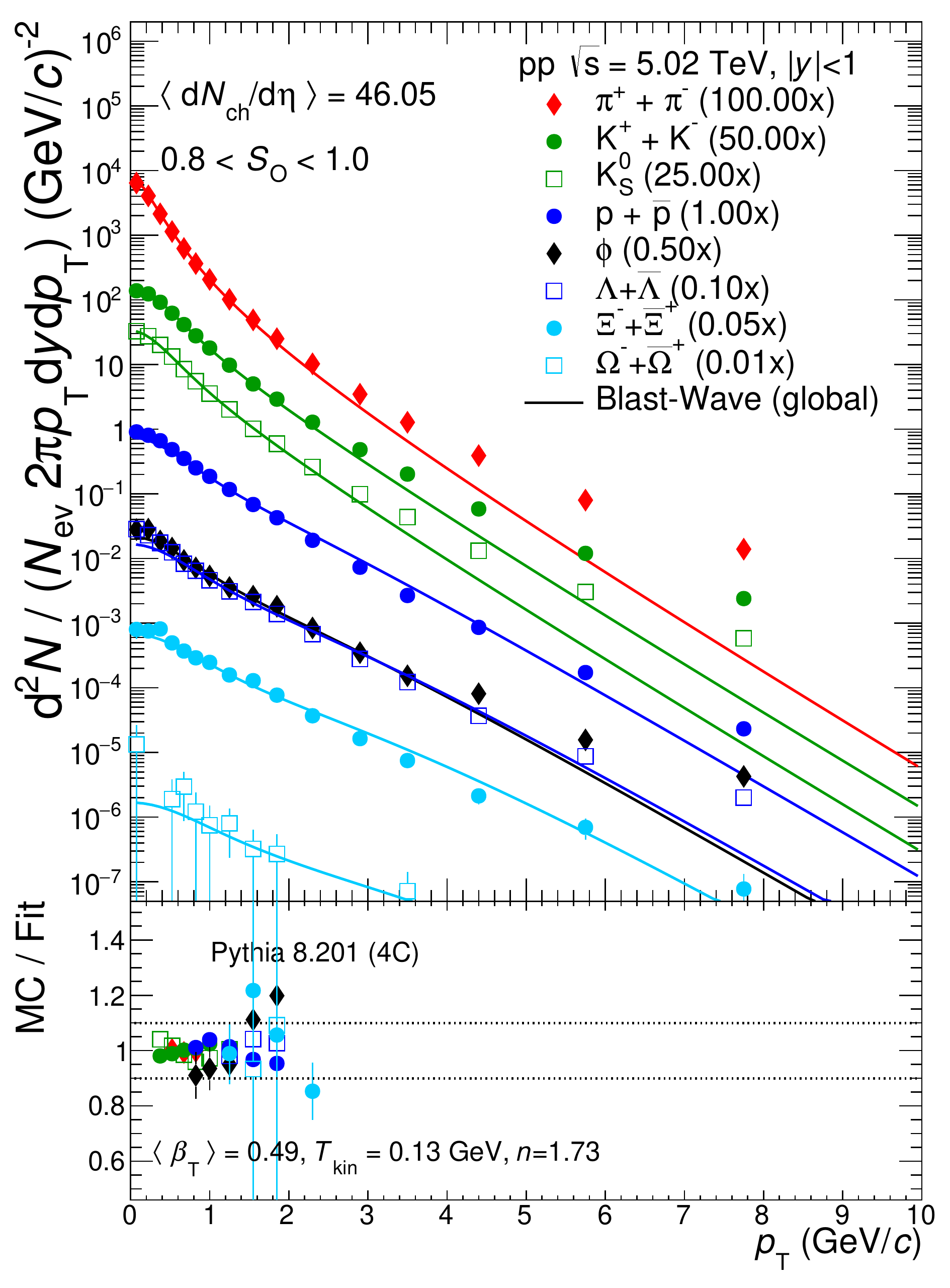}
}
\vspace*{0.5cm}       
\caption{(Color online). Transverse momentum distributions of identified hadrons for jetty-like (upper) and isotropic (bottom) events. Results for three different multiplicity classes are shown. A simultaneous blast-wave fit has been implemented, the results (lines) are plotted together with the \pt spectra.}
\label{fig:5}       
\end{center}
\end{figure*}

The same trend is observed if, instead of jets, inclusive charged hadrons are considered. In this case \meanpt in low transverse spherocity events shows a steeper rise with multiplicity than in the MB case. In contrast \meanpt , for isotropic events shows a weaker multiplicity dependence. This suggests that the choice of narrow \so bins would allow studies with much less fluctuations. 

\subsection{Proton-to-pion ratio  as a function of multiplicity and transverse spherocity}

The particle ratios are among the important observables in nucleus interactions, for example, the proton yield normalized to that for pions encodes flow information and allows to test new hadronization mechanisms like quark recombination/coalescence~\cite{PhysRevC.82.034907}. Again, multiplicity alone is not able to display prominently the important features. The left panel of Fig.~\ref{fig:4} shows that the ratio as a function of \pt is approximately the same for inelastic (minimum bias) and high multiplicity events. In fact, a weak flow-like effect is observed, namely, the depletion of the ratio at low \pt for high multiplicity events with respect to the MB case. For comparison, in the right panel a selection using transverse spherocity is shown,  in this case the ratio for low \so events does not exhibit a ``bump''. On the other hand, for isotropic events (large \nmpi implying stronger flow-like behavior) the ratio displays a similar behavior to that observed in LHC data~\cite{Veldhoen:2012ge}. At low \pt there is a depletion when the multiplicity increases, then  a crossing point is observed, in this case at \pt$\approx$2.5 GeV/$c$. This crossing is followed by an enhancement and, at higher \pt ($>8$ GeV/$c$), the ratio returns to the value obtained for MB. 
This result carries two messages:  the jetty-like events have, if any, a much lower flow-like effect than isotropic events; and the \pt spectra and particle ratios should be extracted for both event classes in order to understand the role of jets when one studies, for example, radial flow. 

\subsection{Results from the blast-wave analysis and comparison with experimental data}

It has been discussed in a previous letter~\cite{Ortiz:2013yxa} that the transverse momentum distributions of identified hadrons obtained from Pythia 8.180 exhibit flow-like features, the effect has been traced to color reconnection and multi-parton interactions. In heavy ion collisions the radial expansion can be extracted through a blast-wave analysis, where, a simultaneous fit to identified hadron \pt distributions for each multiplicity bin is done. This parameterization assumes a locally thermalized medium, expanding collectively with a common velocity field and undergoing an instantaneous common freeze-out~\cite{PhysRevC.48.2462}. The simultaneous fit to all particle species under consideration can provide insight on the common kinetic freeze-out properties of the system. However, one needs to keep in mind that the values which come out from the fit depend substantially on the fit range.

In this section the results from the blast-wave analysis reported by ALICE for \ppb collisions are compared with those obtained from a similar analysis applied to \pp collisions simulated with Pythia 8.180 at  $\sqrt{s}=$5.02 TeV. The parameters extracted from the fits are studied as a function of event multiplicity and transverse spherocity. For the calculation of the transverse spherocity only primary charged hadrons with transverse momenta above 0.15 GeV/$c$ and $|\eta|<1$ are considered. The event multiplicity (\pt spectra) is (are) calculated counting primary charged hadrons within $|\eta|<1$ ($|y|<1$). The transverse momentum distributions of the particle species used in the blast-wave analysis are displayed in Table~\ref{tab:1} along with their corresponding fit ranges. Figure~\ref{fig:5} shows 
some examples of \pt distributions at high multiplicity and for two extreme transverse spherocity event classes: jetty-like ($\so<0.2$) and isotropic ($\so>0.8$). Within 10\% all the MC \pt spectra are well described by the blast-wave model assuming a common transverse velocity $\langle \beta_{\rm T} \rangle$ and temperature $T_{\rm kin}$. Using somewhat different fit ranges for pions, kaons, protons and $\Lambda$; a similar observation is mentioned by the ALICE Collaboration in the analysis of \ppb data~\cite{Abelev:2013haa}.

\begin{table}[b]
\begin{center}
\caption{\label{tab:1} Fit ranges used in the blast-wave analysis.}
\begin{tabular}{cc}
\hline \hline 
\textrm{Particle species} &
\textrm{ Fit range }\\
\hline 
$\pi^{+}+\pi^{-}$ & 0.5-1.0 GeV/$c$ \\
${\rm K^{+}+K^{-}}$ &  0.3-1.5 GeV/$c$ \\
${\rm K_{S}^{0}}$ &  0.3-1.5 GeV/$c$  \\
${\rm p + \bar{p}}$ &  0.8-2.0 GeV/$c$  \\
$\phi$ &  0.8-2.0 GeV/$c$  \\
$\Lambda$ &  1.0-2.1GeV/$c$  \\
$\Xi^{-}+\bar{\Xi}^{+}$ & 1.2-2.6 GeV/$c$  \\
$\Omega^{-}+\bar{\Omega}^{+}$ & 1.3-2.8 GeV/$c$  \\
\hline \hline 
\end{tabular}
\end{center}
\end{table}

\begin{figure}
\begin{center}
\resizebox{0.5\textwidth}{!}{%
  \includegraphics{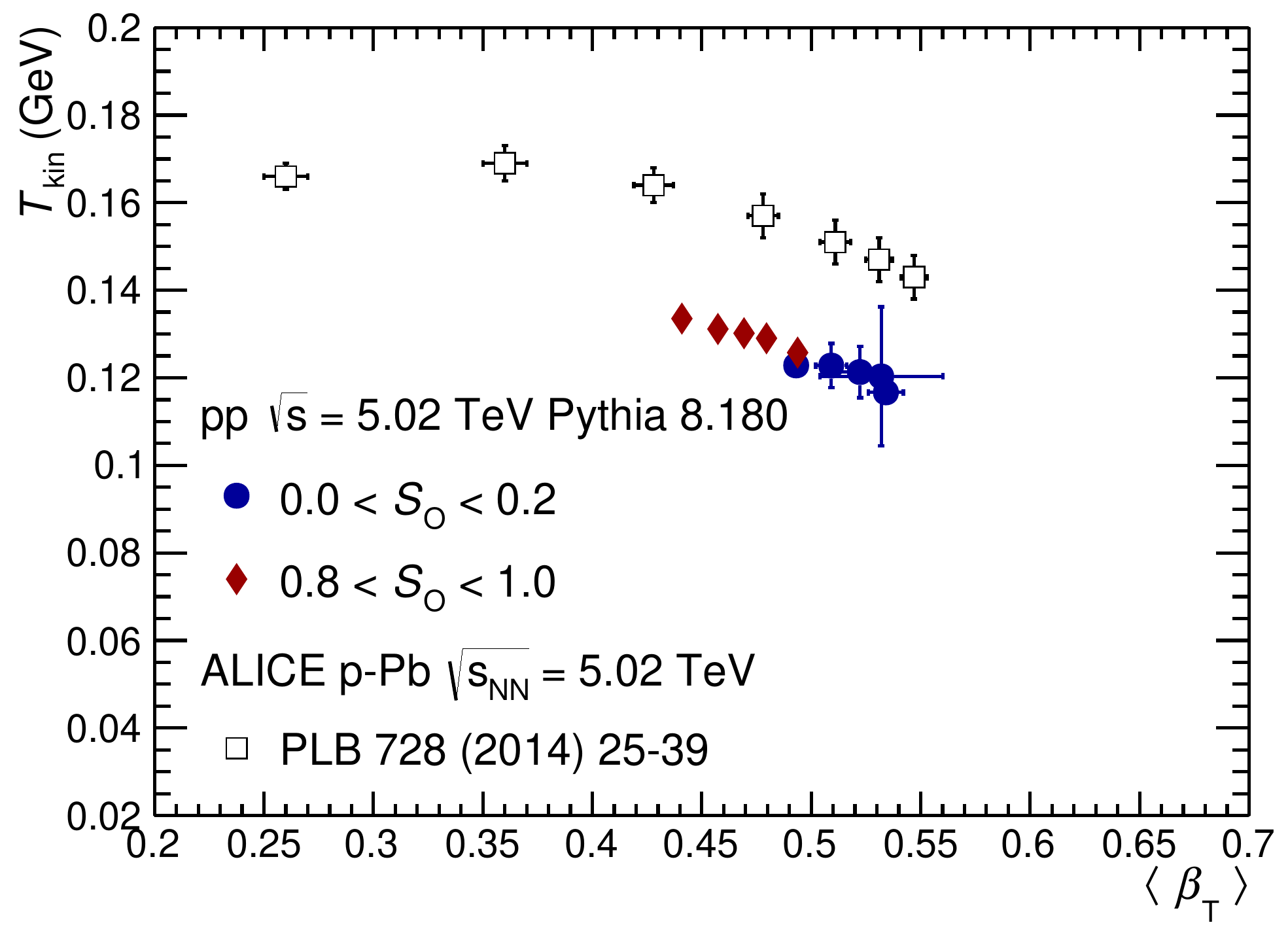}
}
\caption{(Color online). Results from the combined blast-wave fit to the charged pion, kaon and (anti)proton \pt distributions in \pp collisions at $\sqrt{s}$ = 5.02 TeV. The distributions were generated with Pythia 8.180, results are presented for jetty-like events (blue circles) and isotropic events (red circles). MC results are compared with experimental data (black squares).}
\label{fig:6}       
\end{center}
\end{figure}

Finally, Fig.~\ref{fig:6} shows the evolution of  $\langle \beta_{\rm T} \rangle$ vs. $T_{\rm kin}$ with event multiplicity and transverse spherocity. The results are compared with ALICE \ppb data. The low transverse spherocity (jetty-like) events do
exhibit a high $\langle \beta_{\rm T} \rangle$ value with a low $T_{\rm
kin}$, while the high spherocity events exhibit a
 lower $\langle \beta_{\rm T} \rangle$ and higher $T_{\rm kin}$. This represents in
retrospect an evidence:  the jetty-like events by their  nature will
have a tendency to mimic flow, consequently, in radial flow studies
care needs to be taken to avoid the non flow effects. The comparison of the
simulations with    the experiment
 demonstrates that in the treatment of the blast wave analysis one has
 to be careful to identify  effects coming from jets from those from
collective phenomena.

\section{Conclusions}

The transverse spherocity has been used to study  the influence of the multiparton
interactions on the final state in pp collisions. It was demonstrated that the
number of multi-parton interactions is strongly correlated with the final event
multiplicity  at least up to a point where a saturation of the number of
multi-parton interactions occurs. The transverse spherociy selection allows to
identify and analyze two extreme cases: the jetty-like and the isotropic events
situated at the two ends of the transverse spherocity spectrum. In the transverse spherocity event classes, narrower \nmpi distributions are achieved than in the case without any selection on event shape. 
To illustrate the application of these variables, different studies were done, namely,  jet production, identified particle ratios and blast-wave
analysis. The results show the benefits of the combined multiplicity and event shape analysis.
We conclude that a more widespread use of event shape variables in the analysis of
the data may bring us a much better understanding  of the detail of the collisions.
The results using Pythia 8.180 simulations  are qualitatively very similar to the
experimental observations in \ppb and \pbpb data (e.g., different particle composition in the jet and in the bulk regions). Suggesting that more differential studies using event shapes reveal interesting features which could be exploited to get physical information as well as to  improve models used in the MC generators.

\section{Acknowledgments}
The authors acknowledge the technical support of Luciano Diaz. Support for this work has been received by CONACyT
under the Grant No. 260440; and by DGAPA-UNAM under PAPIIT grants IN105113, IN107911, IN108414 and IA102515. The EPLANET funding has facilitated the necessary meetings at CERN.

\bibliographystyle{elsarticle-num}

\bibliography{biblio}

\end{document}